\documentclass[iop,apj,twoside,onecolumn,english]{aastex62}

\usepackage{graphicx}
\usepackage{longtable}
\usepackage{amssymb}
\usepackage{multirow}
\usepackage{rotating}

\usepackage{url}
\usepackage{color}
\definecolor{purple}{rgb}{.9,0,.1}
\usepackage{natbib}
\begin{document}

\journalinfo{to appear in ApJ}

\newcommand{\kms}{km s$^{-1}$}
\newcommand{\msun}{M$_{\sun}~$}
\newcommand{\rsun}{R$_{\sun}~$}
\newcommand{\lsun}{L$_{\sun}~$}
\newcommand{\gbpi}{Gaia\,17bpi}

\title{Gaia\,17bpi: An FU Ori Type Outburst}

\author{Lynne A. Hillenbrand} 
\affiliation{Department of Astronomy, California Institute of Technology, Pasadena CA 91125}
\author{Carlos Contreras Pe\~{n}a} 
\affiliation{School of Physics, University of Exeter, Stocker Road Exeter, Devon, EX4 4QL}
\author[0000-0001-6352-5312]{Sam Morrell}
\affiliation{School of Physics, University of Exeter, Stocker Road Exeter, Devon, EX4 4QL}
\author[0000-0002-0506-8501]{Tim Naylor}
\affiliation{School of Physics, University of Exeter, Stocker Road Exeter, Devon, EX4 4QL}
\author[0000-0002-0631-7514]{Michael A. Kuhn}
\affiliation{Department of Astronomy, California Institute of Technology, Pasadena CA 91125}
\author[0000-0002-0077-2305]{Roc M. Cutri}
\affiliation{IPAC, California Institute of Technology, Pasadena CA 91125}
\author[0000-0001-6381-515X]{Luisa M. Rebull}
\affiliation{IPAC, California Institute of Technology, Pasadena CA 91125}
\author{Simon Hodgkin}
\affiliation{Institute of Astronomy, University of Cambridge, Madingley Road, Cambridge CB3 0HA, UK}
\author{Dirk Froebrich}
\affiliation{Centre for Astrophysics \& Planetary Science, The University of Kent, Canterbury, Kent CT2 7NH, UK}
\author{Amy K. Mainzer}
\affiliation{JPL, 4800 Oak Grove Dr. Pasadena 91109}

\begin{abstract}
We report on the source \gbpi\ and identify it as a new, ongoing FU Ori type outburst,
associated with a young stellar object.  
The optical lightcurve from {\it Gaia} exhibited a 3.5 mag rise with
the source appearing to plateau in mid/late 2018.  
Mid-infrared observations from {\it NEOWISE} also show 
a $>$3 mag rise that occurred in two stages, with the 
second one coincident with the optical brightening, and the 
first one preceding the optical brightening by $\sim$1.5 years. 
We model the outburst as having started between October and December of 2014. 
This wavelength-dependent aspect of young star accretion-driven outbursts 
has never been documented before.  Both the mid-infrared and the optical
colors of the object become bluer as the outburst proceeds.
Optical spectroscopic characteristics in the outburst phase include: 
a GK-type absorption spectrum, strong wind/outflow in 
e.g. Mgb, NaD, H$\alpha$, \ion{K}{1}, \ion{O}{1}, and \ion{Ca}{2} profiles, 
and detection of \ion{Li}{1} 6707 \AA.  
The infrared spectrum in the outburst phase is similar to that of an 
M-type spectrum, notably exhibiting prominent $H_2O$ and  
$^{12}$CO (2-0) bandhead absorption in the K-band, and likely 
\ion{He}{1} wind in the Y-band.
The new FU Ori source \gbpi\ is associated with a little-studied 
dark cloud in the galactic plane, located at a distance of 1.27 kpc.
\end{abstract}


\section{INTRODUCTION}

The accretion paradigm for young stars is reviewed by \cite{hartmann2016}.
It features an early spherical-like infall from a hydrostatic core
of slowly rotating molecular and dusty material, the build-up of (proto)stellar
mass through a combination of the direct collapse and accretion from a disk
(mediated by a fraction of the mass that is lost in accretion-driven 
winds/outflows), and finally, a late-accretion phase that competes with 
planet formation and remnant winds in completely depleting the disk.
The initially high accretion rates are believed to generally decline 
with age. However, our physical understanding of star formation and stellar mass
assembly currently relies on episodic accretion, or punctuated periods 
of enhanced mass accretion/outflow, in order to build up the needed
stellar mass on the required time scales.

Based on a small sample of ``classical" FU Ori stars \citep{herbig1977}
and a few subsequently discovered FU Ori-like objects -- those whose transition
to the outburst state had not been observed -- \cite{HK1996} summarized
the basic scenario of episodic accretion.  They envisioned that enhanced
accretion would occur more frequently in more massive disks at early
protostellar stages, and thus that the rate of FU Ori events would be
higher for protostars, and lower for optically revealed pre-main sequence stars.
Subsequent detailed modeling of instabilities arising in the inner disk 
\citep{bae2014} or the outer disk \citep{vorobyov2015} has produced
quantitative predictions concerning the amplitudes, durations,
and duty cycles of episodic accretion in young stellar objects.

A long-standing problem for the importance placed on the FU Ori scenario
in building up stellar mass, is that FU Ori outbursts are rare.
Over the past seven decades, fewer than 13 actual FU Ori outbursts 
have been recorded, with another $\sim$13 sources identified as 
FU Ori-like based on their present spectra and spectral energy distributions 
obtained in a hypothetical post-outburst state \citep{RA2010,connelley2018}.  
Although the discoveries are increasing in number with time 
\citep[Figure 73 in][]{reipurth2016}, the rate of FU Ori
outbursts remains rather poorly constrained empirically. 
This is an especially notable gap when compared to our state of knowledge 
regarding event rates for e.g. cataclysmic variable and other novae,
supernovae of various classes, and even tidal disruption events.

Modern all-sky and all-hemisphere time domain surveys have the potential 
to better constrain the true FU Ori rate \citep{hillenbrand2015}, 
and better illuminate the diversity of young star outbursts of various types,
e.g. FU Ori-like vs V1647 Ori-like vs EX Lup-like \citep{cp2017a,cp2017b}. 
Only when the full phase space of young stellar object variability 
is more completely mapped out, can we improve our understanding of how stars gain their mass.

The {\it Gaia} mission is one such photometric survey, performing repeated scans
of the sky primarily directed towards establishing accurate and precise astrometry.
However, changes in source brightness trigger alerts that are made publicly available
via a URL interface \footnote{Gaia Alerts;\ {\url{http://gsaweb.ast.cam.ac.uk/alerts}}}.  
These pages are monitored by groups wishing to identify 
objects of interest to their science goals, and perform follow-up observations.
An alert for \gbpi\ was issued on 2017, June 21, following a 2 mag brightness
increase relative to observations from the earliest {\it Gaia} epochs.  
The source has continued to brighten.

While the lightcurve of \gbpi\ exhibits an FU Ori-like rise, several classes of
large-amplitude pulsating variables (LPVs) have similarly slow rise times, 
as can active galactic nuclei (AGN).  However, the location of \gbpi\ 
in proximity to an optically opaque cloud and region of infrared nebulosity, 
suggested that further investigation of the source was warranted.

In this paper we report on the environment of the new outburst source \gbpi\
(\S2), on the characteristics of the pre-outburst object (\S3), 
and 
on the substantial rise relative to previous photometry
exhibited in the {\it Gaia} lightcurve as well as in {\it NEOWISE} observations, 
including follow-up optical photometry (\S4.1-4.3).  In \S4.4 
we present new outburst-era spectroscopy at optical and infrared wavelengths,
and \S5 contains a short discussion and our conclusions.

\section{THE STAR-FORMING REGION CONTAINING THE NEW OUTBURST SOURCE}

\gbpi\ is associated with a previously faint 
optical (r$<$22 mag), near-infrared, and mid-infrared point source.
The position is located towards the northern end of the elongated G53.2 
``infrared dark cloud", just outside of a particularly opaque region on optical and infrared images, and adjacent to a small \ion{H}{2} region cataloged as
\object{HRDS G053.822-00.057} \citep{anderson2014}. 
The modest star-forming region is hardly studied.

The only previous relevant publication covering objects in the vicinity of \gbpi\ 
is by \cite{kim2015}. These authors illustrate the morphology of the dust (based on Bolocam) 
and gas (based on GRS) relative to the mid-infrared emission and absorption (see their Figure 1). 
\gbpi\ appears to be on the periphery of the cloud, beyond the
CO gas contours that also delineate the optically opaque region, 
but between two millimeter-wavelength dust clumps.
\cite{kim2015} used {\it Spitzer} photometry to identify several hundred young stars 
as likely members of the star forming region which extends in a narrow filament
over several degrees to the southeast of \gbpi, 
though not including the newly outbursting source that we report here.
\cite{kim2015} quote a kinematic distance to the cloud of 1.7 kpc.

\begin{figure}
\includegraphics[width=1.00\textwidth]{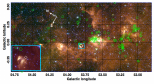}
\caption{
YSO's identified as described in the text,
plotted over a {\it Spitzer} Space Telescope \citep{werner2004} image constructed from
8.0 micron (red), 5.8 micron (green), and 3.6 micron (blue)
data taken as part of the GLIMPSE-I survey \citep{churchwell2009}.
The position of \gbpi\ is indicated by the large diamond; 
inset shows an expanded view $4.5\arcmin \times 4\arcmin$ in size illustrating
the adjacent small \ion{H}{2} region HRDS G053.822-00.057. 
}
\label{fig:region}
\end{figure}

In order to refine the distance and to assess the overall environment of the outburst source,
we repeated the \cite{kim2015} analysis by using the Spitzer/GLIMPSE \citep{churchwell2009}, 2MASS \citep{cutri2003,skrutskie2006},
and Gaia DR2 \citep{gdr2} catalogs.  The color criteria of \cite{gutermuth2009}
were used to identify infrared excess sources that are likely young stellar objects,
then Gaia proper motions and parallaxes were used to down-select to a sample of reliable members. 
From infrared excess alone, there is substantial contamination from dusty giants in the background.
From Gaia alone, it is difficult to identify the association due to an insufficient 
population of stars that would stand out in proper-motion/parallax space. 
As in \cite{kim2015}, \gbpi\ was not selected by our mid-infrared selection 
methods, mainly because in the available {\it Spitzer} catalog data it is detected  
in only the two shortest (3.6 and 4.5 $\mu$m) photometry bands. 
The spatial distribution of our infrared excess and kinematically selected sources 
exhibits two clumps to the southeast of \gbpi, coincident with the morphology of the opacity
in optical/infrared images.  Again, \gbpi\ is found to be located on the periphery of the star forming region.

We assess the cloud distance based on examination of the Gaia parallax distribution 
of the infrared excess sources, using a weighted median approach identical to that described in \cite{kuhn2018}. 
Our value of $1270^{+80}_{-70}$ pc (including both random and systematic error) 
is approximately 25\% closer than the previous kinematic distance estimate.  
The total cloud mass reported by \cite{kim2015}, 
correcting for the distance revision, becomes a few $\times 10^4 M_\odot$.
The cloud is comparable in its filamentary morphology and its size and mass
to the Taurus dust cloud and molecular gas complex \citep{pineda2010}.
The cloud radial velocity reported by \cite{kim2015} is 
$22.9 \pm 1.0$ km/s, which if an LSR value corresponds to $v_{helio}= 4.5 km/s$.

\section{THE PRE-OUTBURST OBJECT}

\subsection{SED Data Collection}

The spectral energy distribution (SED) of the pre-outburst object
was assembled from catalog data originating from the Gaia DR2,
PanSTARRS DR1, IPHaS, 2MASS, and Spitzer/GLIMPSE-I surveys, 
as well as new data reduction of {\it Spitzer} and Herschel images.
The results of the reductions described below are given in 
Table~\ref{tab:sed}.

For the {\it Spitzer} data, the GLIMPSE catalog contains 
measurements only at the shortest wavelengths, 3.6 and 4.5 micron. 
In order to better 
characterize the infrared SED, we downloaded from IPAC/IRSA all {\it Spitzer} image data 
from IRAC (3.6, 4.5, 5.8, 8 micron) and MIPS (24, 70 micron), 
and all Herschel data from PACS (70, 160 microns) and SPIRE (250, 350, 500 microns).
The source is apparent in all four IRAC bands, but at the longer wavelengths
there is contamination from a very bright source to the northeast and only upper limits
could be derived, at much higher levels than typical in less confused regions.

The position was observed at mid-infrared wavelengths as part of the GLIMPSE program (Benjamin et al. 2003; Churchwell et al. 2009) with IRAC in late 2004. 
The mosaics we used were assembled as part of the {\it Spitzer} Enhanced Imaging Products\footnote{SEIP;\ {\url{https://irsa.ipac.caltech.edu/data/SPITZER/Enhanced/SEIP/overview.html}}} which summed up all available data from the cryogenic phase of the mission.  GLIMPSE is a relatively shallow survey, but the target is clearly visible in the IRAC-1 (3.6 um) and IRAC-2 (4.5 um) bands; it is faint but present in IRAC-3 (5.8 um) and IRAC-4 (8 um).   Aperture photometry was performed at the location, using an aperture of 3 native pix, with an annulus of 3-7 native pix. (A native pixel is 1.2$\arcsec$.) Aperture corrections of 1.124, 1.127, 1.143, and 1.234 were used, as described in the IRAC Instrument Handbook\footnote{SSC;\ {\url{http://irsa.ipac.caltech.edu/data/SPITZER/docs/irac/iracinstrumenthandbook/}}}. Reported errors are statistical and do not include systematic effects.  

This region was also observed by MIPS at 24 microns as part of the MIPSGAL program \citep{carey2009} in late 2005, and mosaics re-generated as part of SEIP.  However, the source was not detected. Calculating photometric limits at this location is complicated by nearby bright ISM; we performed aperture photometry at the location as if there was a source there, following suggestions from the MIPS Instrument Handbook\footnote{SSC;\ {\url{http://irsa.ipac.caltech.edu/data/SPITZER/docs/mips/mipsinstrumenthandbook/}}} that is, 5.6$\arcsec$ aperture, annulus 5.6-10.4$\arcsec$, and an aperture correction of 2.05. 

This region was also observed by Herschel \citep{pilbratt2010} in late 2011 as part of the Hi-GAL program \citep{molinari2010} 
using PACS 
and SPIRE. 
The source is not visible in the Herschel High-Level Images\footnote{HHLI;\ {\url{http://irsa.ipac.caltech.edu/data/Herschel/HHLI/overview.html}}}, which sum up all available data. As with MIPS-24, limits are not terribly constraining because of the bright ISM near the region. Aperture photometry obtained from the corresponding instrument handbook was performed at the location to estimate upper limits in each available band (PACS: 70 and 160 um; SPIRE: 250, 350, and 500 um).

\begin{table}
\begin{centering}
\caption{Newly Derived Infrared Photometry\tablenotemark{a} for the \textit{Gaia} 17bpi Progenitor}
\begin{tabular}{| c | c | c | c |}
\hline
Instrument & Wavelength ($\mu$m) & Flux (Jy) & Error (Jy) \\
\hline
Spitzer/IRAC & 3.6  &312.26e-6& 43.09e-6   \\ 
Spitzer/IRAC & 4.5  &293.32e-6& 43.34e-6   \\ 
Spitzer/IRAC & 5.8  &499.99e-6& 142.61e-6 \\ 
Spitzer/IRAC & 8.0  &754.89e-6& 247.29e-6   \\ 
Spitzer/MIPS & 24   &$<$0.058 &\nodata   \\ 
Herschel/PACS& 70   &$<$1.14  &\nodata  \\
Herschel/PACS& 160  &$<$3.92  &\nodata  \\
Herschel/SPIRE& 250  &$<$3.24  &\nodata  \\
Herschel/SPIRE& 350  &$<$2.61  &\nodata  \\
Herschel/SPIRE& 500  &$<$2.76  &\nodata \\  
\hline
\tablenote{Observations were taken in 2004 as part of the {\it Spitzer} GLIMPSE and MIPSGAL programs.}
\end{tabular}
\label{tab:sed}
\end{centering}
\end{table}

\subsection{SED Analysis}

\begin{figure}
\includegraphics[width=0.50\textwidth]{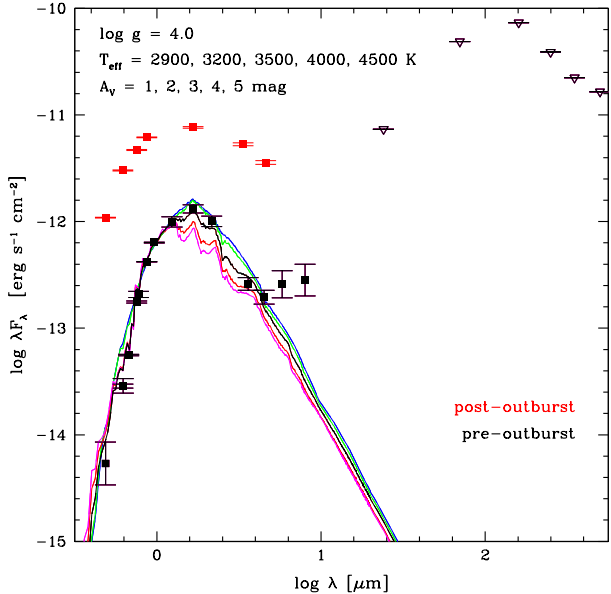}
\vskip-1truein
\caption{
Spectral energy distribution of \gbpi. 
Red points are the post-outburst color data presented here. 
Black points are the assembled pre-outburst photometry 
with black downward triangles indicating upper limits.  
Also shown are
reddened model photospheres normalized at z-band to the pre-outburst photometry. 
See text for sources of data and models.
The empirical SED in the pre-outburst state is consistent with 
a photospheric temperature in the range of 2900-4500 K, with more reddening
required for the hotter temperatures, and an infrared excess.
The excess is clearly implied by the 8.0 and 5.8 micron measurements,
but at 4.5 micron and shorter wavelengths it is dependent 
on the true photospheric temperature.
Note that the pre-outburst sets of grizy, JHK, and IRAC1,2,3,4 measurements 
were not obtained contemporaneously, and thus the compiled SED could be deceptive.
}
\label{fig:sed}
\end{figure}

The 0.5-4.5 micron photometric measurements and the longer wavelength 
upper limits are shown in Figure~\ref{fig:sed}.  It may be important to note
that the pre-outburst photometry consists of measurements taken at different
times, with the optical grizy data taken in 2012, the
near-infrared JHK data taken in 1999, and the IRAC1,2,3,4 measurements 
taken in 2004.  As these sets are non-contemporaneous, the compiled SED 
could be deceptive.

Photospheric models ranging in temperature from 2900-3200-3500-4000-4500 K
from NextGen2 \citep[following][]{hauschildt1999} result in acceptable fits to the Wien part of the SED 
for corresponding visual extinction values of 1-2-3-4-5 mag, respectively. 
Photospheres warmer than $\sim5500$ K can not be reddened enough 
to match the bluer wavelengths without exceeding the measurements around and 
beyond the SED peak.  Photospheres cooler than $\sim2700$ K do not
match at the bluer wavelengths, even unreddened.

Hotter temperatures minimize the near-infrared excess;
however, these warmer temperatures
do not match the r-band to G-band to i-band data as well as cooler temperatures do.
Temperatures cooler than $\sim4000$ K imply a near-infrared excess, the
amplitude of which increases as the assumed photospheric temperature decreases.
The 5.8 and 8.0 micron points clearly exceed the expectations from a reddened 
photosphere of any temperature.  Our preferred model, derived by considering 
a number of different normalization schemes, is the 3500 K photosphere with $A_V=3$ mag.

\subsection{Other Attributes}

Unfortunately we know little else about the pre-outburst object
beyond its location and its colors.
It was too faint for parallax or proper motion reporting 
in Gaia DR2, for example.

Integrating the pre-outburst photometry described above 
between 0.45 and 4.5 microns, and assuming the 1.27 kpc distance derived above,
a source luminosity of 0.3 \lsun is derived for the pre-outburst object.  
The true value will be slightly higher once the proper reddening correction 
is determined.  This luminosity is seemingly appropriate for a 
pre-main sequence star in the above temperature range.
The corresponding spectral type would be mid-K to M.

In the {\it Gaia} photometry presented below, there is some indication
of photometric variability in the earliest G-band measurements,
taken pre-outburst.  \gbpi\ is quoted in the Gaia DR2 catalog at
$G = 20.44 \pm 0.015$ mag.  Given that 13 transits contribute to that 
measurement, the individual measurements are implied to have uncertainty 
0.05 mag.  However, as the uncertainty is based on the scatter 
between data points, any intrinsic variability will be included in the  
uncertainty value. Indeed most sources at this magnitude in the general area
have an uncertainty of only 0.005 mag. The possible pre-outburst variability 
of \gbpi\ has an amplitude of a few tenths of a magnitude,
and occurs on a timescale of a few days (Figure~\ref{fig:lcs}), 
consistent with ``typical" T Tauri star variability \citep[e.g.][]{cody2018,rebull2018}.

The iPHaS catalog \citep{barentsen2014} provides $r-H\alpha$ color of
$0.88 \pm 0.29$ mag, but the very red $r-i$ color of
$2.63 \pm 0.18$ mag does not make the source an obvious $H\alpha$ emitter,
according to the models in Figure 2 of \cite{barentsen2013}.

The field of \gbpi\ was also covered as part of the UWISH2 survey \citep{froebrich2011}.
Consistent with results presented below, the pre-outburst source is
detected in continuum J-band and K-band images (taken 2006-07-11 for the UKIDSS 
Galactic Plane Survey) but does not show an excess in an $H_2 -K$ 
difference image ($H_2$ data taken 2010-11-26).  
There is thus no signature of a pre-existing large-scale
shocked outflow from the \gbpi\ progenitor, only some jets
in the dark filament to the northwest of the source that are not related,
and some nebulosity to the northeast that is coincident with the mid-infrared
nebulosity illustrated in Figure~\ref{fig:neoimage}.  If an outflow
appears over the next few years, it would be direct evidence for
FU Ori outburst triggering of mass ejection.

\section{THE OUTBURST OF \gbpi}

\subsection{DISCOVERY}

At position 19:31:5.590 +18:27:52.27 (J2000.),
\gbpi\ was flagged by an automated alerts system \citep[][Hodgkin et al. 2019, in preparation]{hodgkin2013} and announced on the Gaia Alerts public feed\footnote{\url{http://gsaweb.ast.cam.ac.uk/alerts/alert/Gaia17bpi/}}
on 2017, June 23. The alert was triggered as a ``delta-magnitude detection" based on the source brightening 
relative to earlier {\it Gaia} photometry by $>1$ mag in two consecutive transits of the satellite over the
source position (in March and then June, in this case).
Among the stream of Gaia Alerts, this particular source was noted as being a potential young star outburst via an ongoing program to identify young stellar objects that are published on the Gaia Alerts feed.  Contreras Pe\~{n}a et al. (2018, in preparation) present a full description of this program. 

Although \gbpi\ was not successfully cross-matched with any specific object in the search catalogs, it was
identified using what we term a ``vicinity match," that is, as being located within 2 arcmin of a confirmed or candidate young stellar object. This methodology leverages the fact that YSOs are located in spatially coherent regions of star formation, hence clustered on the sky. Further, it acknowledges that the current census information for many nearby star-forming regions remains incomplete,
and uses photometric variability to flag potential YSOs that are previously uncataloged. In fact, \gbpi\ has five
stars falling in the vicinity of the alert coordinates, given in \autoref{tab:proxmatch-objects}. 

\begin{table}
\begin{centering}
\caption{SIMBAD-identified YSOs within 2 arcmin of \textit{Gaia} 17bpi}
\begin{tabular}{| c | c | c | c |}
\hline
Name & RA (J2000) & Dec (J2000) & Separation (\arcsec)\\
\hline
2MASS J19310037+1827424 & 19:31:00.38 & +18:27:42.4 & 74.8\\
SSTGLMC G053.8210-00.0303 & 19:31:02.28 & +18:29:14.0 & 94.3\\
SSTGLMC G053.8129-00.0785 & 19:31:12.00 & +18:27:25.0 & 95.2\\
2MASS J19310700+1829328 & 19:31:07.01 & +18:29:32.8 & 102\\
2MASS J19305884+1829007 & 19:30:58.84 & +18:29:00.7 & 117\\
\hline
\end{tabular}
\label{tab:proxmatch-objects}
\end{centering}
\end{table}

\subsection{ASSEMBLY OF LIGHTCURVE DATA}

\begin{figure}
\includegraphics[width=0.8\textwidth,angle=-90]{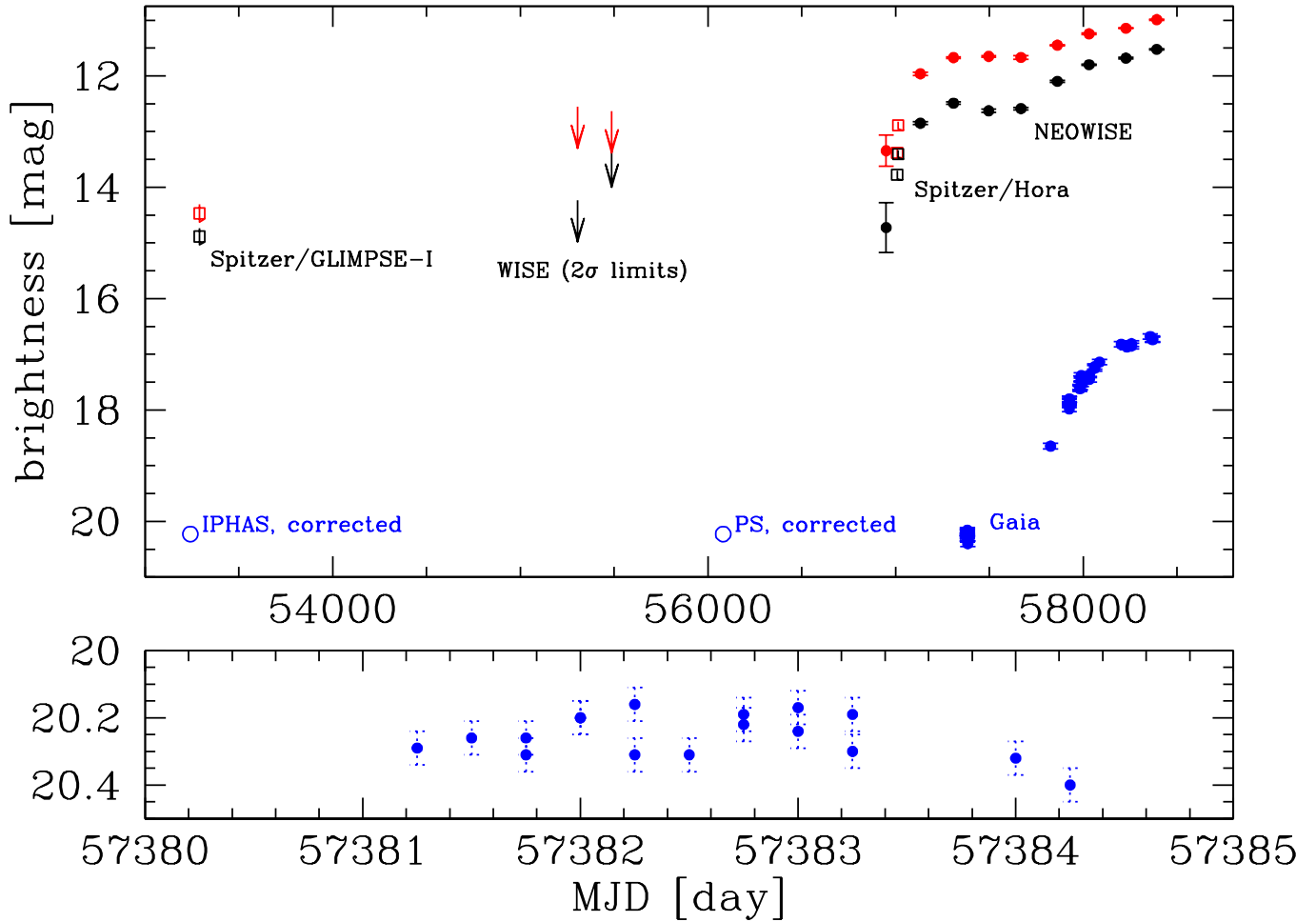}
\caption{
Top:
Lightcurves measured by Gaia (G-band with effective wavelength 0.673 micron)
in blue filled symbols and {\it NEOWISE} (3.4 and 4.6 microns) in black and red
filled symbols.
The recent photometry measuring the outburst is supplemented by previous
data (open symbols) from the IPHaS and PanSTARRS surveys in the optical, 
corrected from r-band to G-band as described in the text, 
and from $Spitzer$ in the infrared, plotted as the native 3.6 and 4.5 micron 
measurements, without correction.
Error bars are shown on all points.
Downward pointing arrows indicate the epochs of position coverage by 
the {\it WISE} sky survey, in which the source was not detected.
Bottom:
Zoom-in to the first epochs of Gaia data, taken in the last days of 2015. 
Notional 5\% uncertainties are plotted as dotted error bars, 
derived by inflating the Gaia DR2 uncertainty for the number of transits
included in the scatter measurement.  If the real uncertainties are 
at this level or smaller}, variability at the several tenths 
of a magnitude level could have been occurring 
before the major brightening episode.
\label{fig:lcs}
\end{figure}

\begin{figure}
\includegraphics[width=0.5\textwidth]{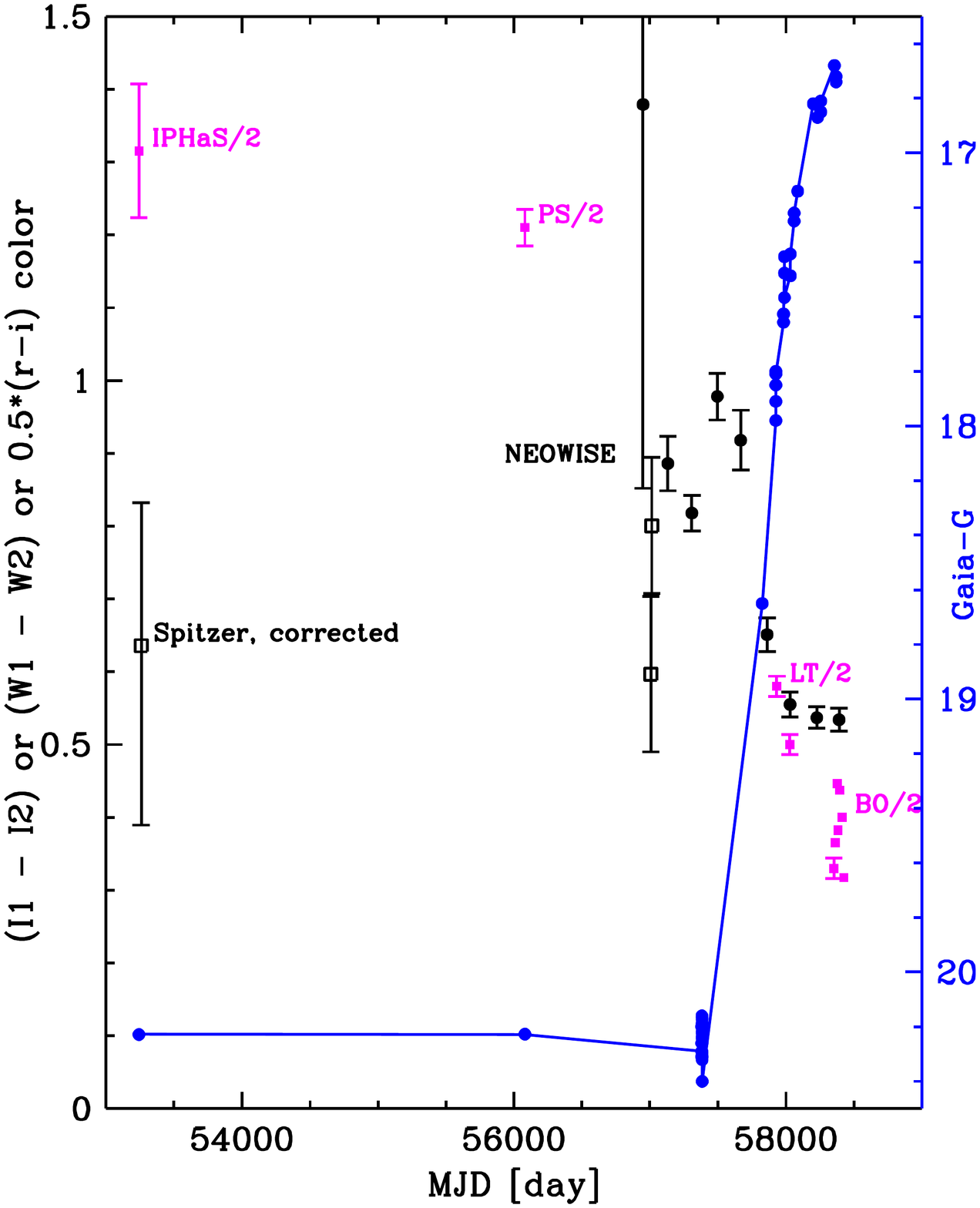}
\caption{
Magenta points illustrate the optical color evolution of \gbpi, 
as measured by the post-outburst photometry presented here 
compared to the pre-outburst IPHaS and PanSTARRS optical color
(corrected as necessary from AB to Vega magnitudes).
The r-i color becomes bluer during the outburst. The optical color change
is substantially larger than indicated by the axis labelling, as the magenta data points 
have been divided by 2 for purposes of easy comparison to the infrared color,
shown in black.
Black points illustrate the infrared color evolution of \gbpi,
as measured by {\it NEOWISE} and {\it Spitzer}, the latter photometry
color-corrected to the NEOWISE photometric system; see text.
The [3.4]-[4.6] color in outburst may be initially slightly redder than the pre-outburst color.  As \gbpi\ becomes optically bright (blue points, same as in Figure~\ref{fig:lcs}), 
the [3.4]-[4.6] color clearly turns bluer, in concert with the optical color.
}
\label{fig:color}
\end{figure}

\subsubsection{Optical}
Photometry from the Gaia mission was downloaded from the alerts service webpage,
with the last update for our analysis occurring on 10 September, 2018.
Only Gaia G-band measurements are available and error estimates are not included.
We do not reproduce the numbers here, as the photometric quality should 
continue to improve as the mission proceeds.

The recent Gaia lightcurve was supplemented by heritage catalog data
from IPHaS \citep{barentsen2014}  
and PanSTARRS \citep{flewelling2016}, Reported r-band magnitudes
were adjusted (in the case of IPHaS, first from Vega to the AB system) 
from the PanSTARRS AB system to Gaia's G-band,
using an empirical calibration\footnote{
We derived $G_{calculated, Vega} = -2.375 + 1.401\times r_{PS, AB} - 0.01694\times r^2_{PS, AB}$ over the magnitude range $13 < G < 22$.
}.
Looking to other sources of recent photometry, we discovered that
although there is PTF/iPTF imaging in the region spanning the time period 
after the PanSTARRS data point, and before/after the first Gaia detections, 
the source sits in a PTF chip gap.  There is thus unfortunately  
no optical photometry recording the initial rise of \gbpi. 

After the 2017 Gaia alert was issued, we monitored the object using 
the Liverpool Telescope. 
Three nights of griz photometry were obtained on 2017, June 28, 
October 2, and 2018, August 22. A single epoch of H-band photometry 
was obtained on 2017, June 28. The griz magnitudes were calibrated  
using Pan-STARRS AB magnitudes for a set of stars around 5 arcmin from 
\gbpi. The H-band data was calibrated in a similar way using 2MASS magnitudes. 
At the time of the first night of monitoring, the object was already 
2 magnitudes brighter in both the optical and the infrared
than the quiescent state as measured in earlier PanSTARRS and 2MASS data.
The measurements are given in \autoref{tab:optphot}. 

Beginning in spring of 2018, 
as \gbpi\ began to show signs of approaching peak brightness, 
the Zwicky Transient Facility (ZTF) became operational.
Complementing the Gaia lighturve, 
there are several tens of  
epochs of g-band and r-band photometry from ZTF
taken between 2018 March through November (when the source set for the season).  
These data will become available with the release of the NSF/MSIP 
public survey part of ZTF.
However, there is a single ZTF measurement from a 
``reference image" with $g=18.505$ and $r=16.970$ mag reported at IPAC/IRSA 
for a date that can not be determined at present. 

Additional optical imaging was performed 
with the University of Kent's Beacon Observatory \citep{froebrich2018}
The telescope is a 17-inch {\em Planewave} Corrected Dall-Kirkham (CDK) 
Astrograph 
with a 4k\,$\times$\,4k Peltier-cooled CCD camera and a B, V, R, I, H$\alpha$ filter set.
The final images for each night in each of the broadband filters 
were combined from separate integrations. 
Standard data reduction using bias, darks and sky-flats have been applied.
BVri photometric measurements are given in \autoref{tab:optphot}. 

\begin{table}
\begin{centering}
\caption{Optical Photometric Follow-up During the Outburst of \textit{Gaia} 17bpi}
\begin{tabular}{| c | l | l | l | c |c|}
\hline
Telescope &MJD& Magnitude& Error& Filter& photometric system \\
\hline
Liverpool& 57933.091466&20.12&0.02&g&PanSTARRS (AB)\\
Liverpool& 57933.097860&18.47&0.01&r&PanSTARRS (AB)\\
Liverpool& 57933.100280&17.52&0.01&i&PanSTARRS (AB)\\
Liverpool& 57933.100863&16.94&0.01&z&PanSTARRS (AB)\\
Liverpool& 58028.982488&18.97&0.02&g&PanSTARRS (AB)\\
Liverpool& 58028.986326&17.55&0.01&r&PanSTARRS (AB)\\
Liverpool& 58028.987833&16.76&0.01&i&PanSTARRS (AB)\\
Liverpool& 58028.988410&16.32&0.01&z&PanSTARRS (AB)\\
Liverpool& 58353.050848&18.28&0.01&g&PanSTARRS (AB)\\
Liverpool& 58353.054681&16.91&0.01&r&PanSTARRS (AB)\\
Liverpool& 58353.056184&16.22&0.01&i&PanSTARRS (AB)\\
Liverpool& 58353.056767&15.77&0.01&z&PanSTARRS (AB)\\
\hline
Liverpool&57933.104537&13.48&0.02&H&2MASS\\
\hline
Beacon & 58362.43&17.474&0.067&V&APASS \\
Beacon & 58362.43&16.795&0.087&r&APASS \\
Beacon & 58362.43&16.063&0.097&i&APASS \\
Beacon & 58379.37&17.567&0.062&V&APASS \\
Beacon & 58379.36&16.944&0.085&r&APASS \\
Beacon & 58379.36&16.051&0.108&i&APASS \\
Beacon & 58383.35&17.575&0.060&V&APASS \\
Beacon & 58383.35&16.783&0.083&r&APASS \\
Beacon & 58383.36&16.027&0.099&i&APASS \\
Beacon & 58395.37&19.254 & 0.135  &B &APASS \\
Beacon & 58395.36&17.667 & 0.062  &V &APASS \\
Beacon & 58395.36&17.024 & 0.084  &r &APASS \\
Beacon & 58395.36&16.149 & 0.096  &i &APASS \\
Beacon & 58412.31& 17.651& 0.070& V & APASS \\
Beacon & 58412.31& 16.902& 0.078& r & APASS \\
Beacon & 58412.31& 16.101& 0.097& i & APASS \\
Beacon & 58425.29& 17.420 & 0.063 &V & APASS \\
Beacon & 58425.29& 16.707 & 0.079 &r & APASS \\
Beacon & 58425.29& 16.072 & 0.097 &i & APASS \\
Beacon & 58452.74 & 17.226 & 0.0622 & V & APASS \\
Beacon & 58452.74 & 16.710 & 0.0776 & r & APASS \\
Beacon & 58452.74 & 15.912 & 0.0959 & i & APASS \\

\hline
\end{tabular}
\label{tab:optphot}
\end{centering}
\end{table}

\subsubsection{Infrared}

\begin{figure}
\includegraphics[width=0.90\textwidth]{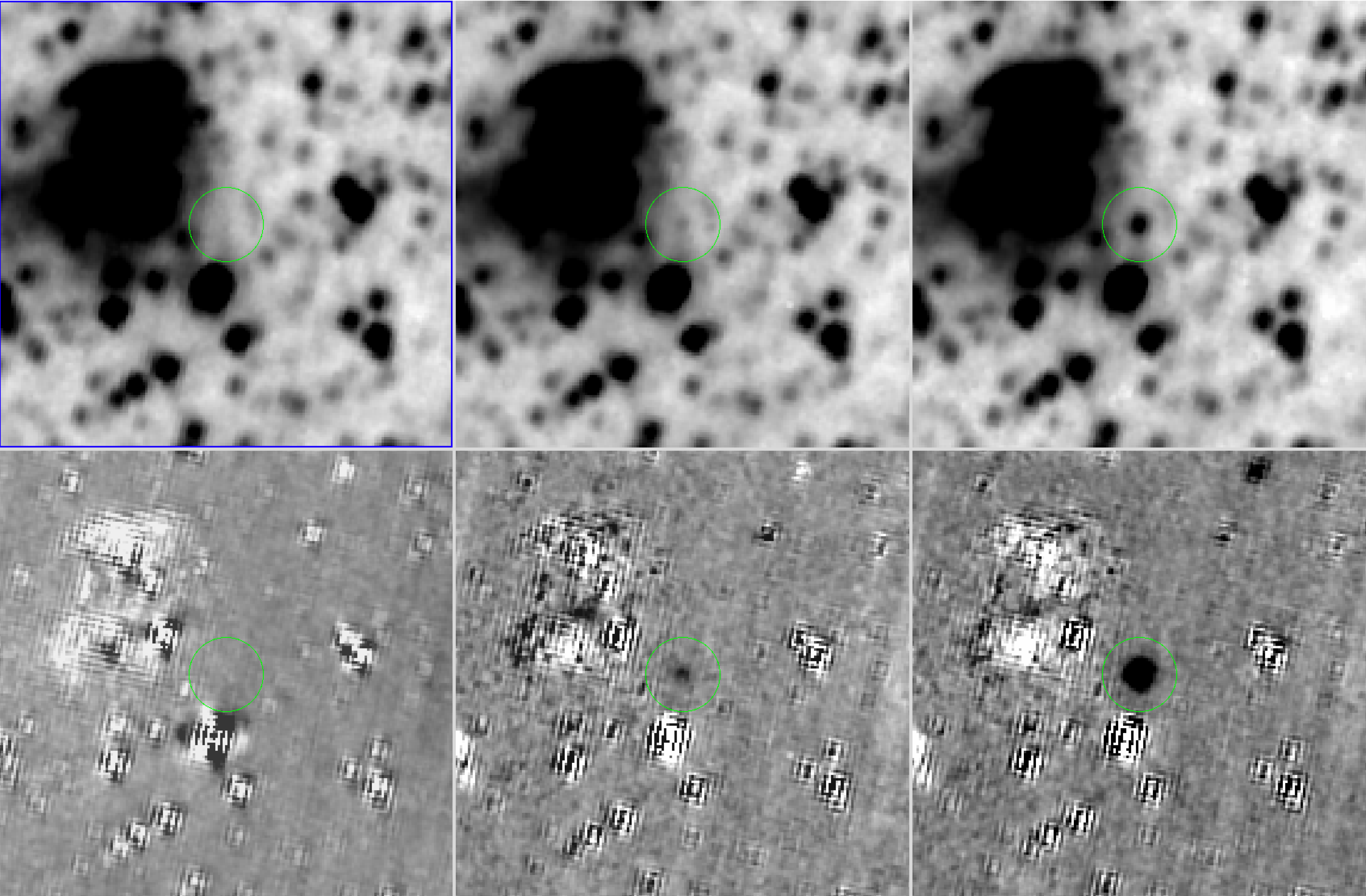}
\caption{
Top row shows the {\it NEOWISE} direct coadded 4.6 micron images from 
(left to right) October 2010, October 2014 and October 2017.
Bottom row shows the difference images relative to an earlier epoch,
specifically (left to right), between October 2010 and April 2010,
 between October 2014 and October 2010, and between October 2017 and October 2010.
 Fields-of-view are $3\arcmin \times 3\arcmin$; green circles
 are $30 \arcsec$ in diameter, centered on the position of the source.
 North is towards the top and east is towards the left.
}
\label{fig:neoimage}
\end{figure}

The position of Gaia 17bpi was observed by {\it WISE} \citep{wright2010} during its primary mission in 2010 April and October, and again twice per year between 2014 October and 2018 October
during the reactivated {\it NEOWISE} survey \citep{mainzer2014}.  {\it NEOWISE} did not observe the source in 2014 April because it lies in the region of sky that was missed due to a short safe-hold during that time \citep{cutri2015}.  

\gbpi\ was not detected in the {\it WISE} observations, and thus it has no entry in the AllWISE Source Catalog.
Brightness upper limits for the 2010 April and October observations 
were estimated using an 8 arcsec aperture 
on coadds of the 15 individual exposures covering the object in each epoch.  
The magnitude upper limits were computed by adding two times 
the flux uncertainty to the measured flux in the aperture.

The source is, however, well-detected at 3.4 and 4.6 micron in the {\it NEOWISE} observations beginning in 2015, April. 
The mid-infrared brightening is illustrated in Figure~\ref{fig:neoimage}.

Because the field is complicated in the mid-infrared, 
due to both nebulosity and to very red nearby bright sources,
the {\it NEOWISE} photometry was validated by examining individual images.
Profile-fit photometric measurements from each individual exposure are
available in the NEOWISE Source Database\footnote{NEOWISE;\ {\url{http://wise2.ipac.caltech.edu/docs/release/neowise/expsup}}} \citep{cutri2015}.  
For the {\it NEOWISE} 2015 April and later epochs, the magnitudes and uncertainties listed in Table~\ref{tab:irphot} were computed from the mean and standard deviation of the mean of the Source Database profile-fit fluxes from all exposures taken during each epoch.  

For the 2014, October observation set, the NEOWISE Source Database has only one
reported detection with an unusually large positional offset relative 
to the later epochs.  The source is not obvious in the individual 
2014, October exposures, and we deemed the one measurement to be spurious. 
However the source is faintly visible at the correct position
in a coadded image formed by combining the 15 individual exposures covering 
the position, as shown in Figure~\ref{fig:neoimage}.  Because of the complexity of
the surrounding field, 3.4 and 4.6 micron photometry
was performed on difference images that were constructed by subtracting coadded 
2010 October images (where the source is not detected) from the coadded 
2014 October images to suppress the nearby confusing objects.  
The flux of the source was measured in an 8 arcsec radius aperture 
on the difference images, which is the same sized aperture used for the 
``standard aperture" photometry in {\it NEOWISE} automated data processing.  
The aperture measurements on the difference images were put on the 
photometric scale of the automated profile-fit measurements from the later 
epochs by applying the same aperture corrections and normalization used 
in the automated processing (these are the w1mcor and w2mcor parameters 
in the {\it NEOWISE} Source Database entries).  This procedure was verified 
by generating difference images for the later NEOWISE epochs 
when the source was well detected.  The calibrated aperture measurements 
on the difference images agree with the average profile-fit measurements 
to within 11\% in W1 and 6\% in W2 in each epoch 
(excluding 2014 October when the source was too faint to be reliably
detected in the individual exposures).
The results appear in Table~\ref{tab:irphot}.

The assembled lightcurve includes the {\it Spitzer}/GLIMPSE-I 
photometry from 2004 discussed above.
In addition, there is a second set of {\it Spitzer} data, taken in 2014, December,   
just after the first {\it NEOWISE} measurements from 2014, October.
The nearby infrared dark cloud
was targetted twice with IRAC 3.6 and 4.5 micron imaging,
observed as part of program 10012 (PI J. Hora). 
Each of these observations are much deeper than GLIMPSE, 
with integration time $\sim$150 sec as compared to $\sim$4 sec, 
and the source was brighter by this time as well.  Aperture photometry was performed as 
described earlier for the GLIMPSE re-reduction, but using aperture corrections of 
1.125 and 1.120, as appropriate for post-cryogenic {\it Spitzer} data. 
The results appear in Table~\ref{tab:irphot}. 

The two epochs of 2014 {\it Spitzer} data appear to have captured
the outburst in progress, with the second epoch brighter than the first (by 0.4 mag at 3.6 micron and 
0.5 mag at 4.5 micron), which we confirmed from visual inspection and repeated photometric measurement.
The only slightly earlier 2014 {\it NEOWISE} measurement is fainter, and the first 2015 {\it NEOWISE}
measurement is brighter (see Figure~\ref{fig:lcs}). 

\begin{table}
\begin{centering}
\caption{Newly Measured Infrared Photometry Before and During the Outburst of \textit{Gaia} 17bpi}
\begin{tabular}{| c | l | l | l | l | l| l|}
\hline
Telescope &MJD& 3.4 $\mu$m Magnitude& Error& 4.6 $\mu$m Magnitude & Error\\
\hline
 $WISE$& 55305.20 &    $>$14.23& 2$\sigma$ limit  &$>$12.55&2$\sigma$ limit \\		
 $WISE$& 55487.10 &	$>$13.25& 2$\sigma$ limit  &$>$12.63&2$\sigma$ limit 	 \\
$NEOWISE$ & 56949.30 &	14.72	& 0.45 &13.35& 0.28		 \\
$Spitzer$ & 57008.16 &13.78 (3.6 $\mu$m)&0.08    &13.38  (4.5 $\mu$m)& 0.08\\
$Spitzer$ & 57014.19 &13.40 (3.6 $\mu$m)&0.07    &12.89  (4.5 $\mu$m)& 0.07\\
$NEOWISE$ & 57131.80 &	12.85	&0.03 &11.96&0.03		 \\
$NEOWISE$ & 57308.40 &	12.49	&0.02 &11.67&0.01		 \\
$NEOWISE$ & 57495.90 &	12.63	&0.03 &11.65&0.01		 \\
$NEOWISE$ & 57667.70 &	12.59	&0.02 &11.67&0.03		 \\
$NEOWISE$ & 57861.90 &	12.10 	&0.02 &11.45&0.01		 \\
$NEOWISE$ & 58030.80 &	11.80 	&0.01 &11.24&0.01		 \\
$NEOWISE$ & 58227.20 &	11.68	&0.01 &11.15&0.01		 \\
$NEOWISE$ & 58391.40 &    11.52  &0.01 &10.99 &0.01             \\
\hline
\end{tabular}
\label{tab:irphot}
\end{centering}
\end{table}

\subsection{QUANTITATIVE ANALYSIS OF THE LIGHTCURVES AND COLOR CURVES}

Figure~\ref{fig:lcs} shows the assembled optical and infrared lightcurve data.
Unfortunately, the beginning of the turn-up from quiescence to an early 
outburst stage is not well-observed in either the optical or the infrared.  
However, the later parts of the optical rise and the optical peak are 
adequately sampled, even at the low cadence of Gaia.  
As of 2018, September, \gbpi\ is $\sim$3.5 mag brighter in the optical 
and $\sim$3 mag brighter in the mid-infrared than it was 5-15 years ago.  

The first Gaia measurements in late 2015  indicate no change in brightness from the historical optical photometry.   
However, about a year before, in late 2014 
there is a pair of {\it Spitzer} measurements that are already 1.1 mag brighter in both I1 and I2
than the {\it Spitzer} I1 and I2 photometry from 10 years previous, in 2004.
The first {\it NEOWISE} points 
were taken just before the two {\it Spitzer} measurements and 1.2 years earlier than the first {\it Gaia} measurement. 
They have large errors, and while the 3.4 micron point is not significantly brighter than the earlier {\it Spitzer} measurement,
the 4.6 micron measurement could be.  
The outburst may have begun between October and December of 2014.

\begin{figure}
\includegraphics[width=0.5\textwidth]{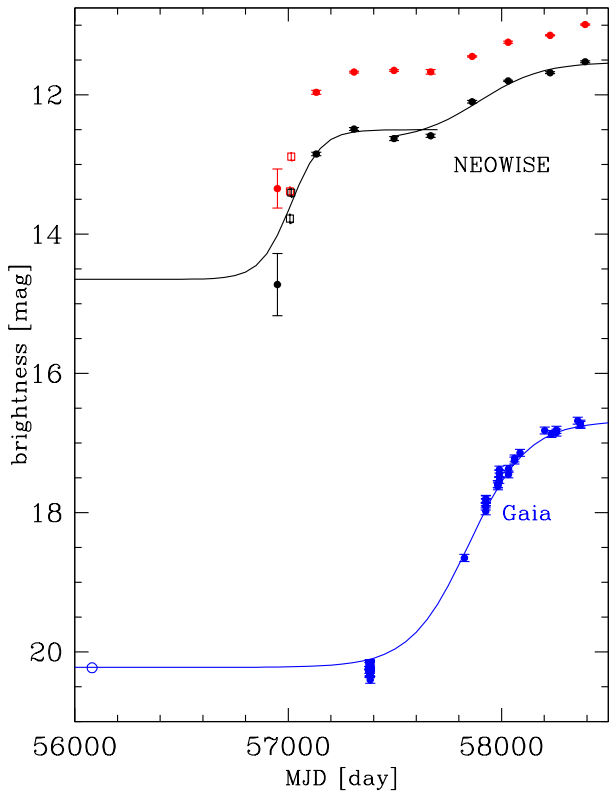}
\caption{
Analytic fits to the FU Ori type outburst of \gbpi.
The $Gaia$ optical lightcurve is fit with a single sigmoid function while the
$NEOWISE$ mid-infrared lightcurve is modeled as having two stages,
described by two sigmoids. See text for fit parameters.
}
\label{fig:fit}
\end{figure}

The evidence from the time series suggests that the outburst may have started in the infrared,
and manifest later in the optical.  As we do not sample the beginning of the burst 
in the optical, given the large time gap between the initial and subsequent Gaia photometry, 
we can not calculate the time delay, but $\sim$1.5-2 years is implied.
The {\it NEOWISE} lightcurve further indicates a possible two-stage rise,
in which there is perhaps 18 months of measurable mid-infrared brightening, followed by 
a plateau of about a year, then another at least 24 months of mid-infrared brightening 
corresponding to the optical rise that is well-sampled by the Gaia measurements.
Figure~\ref{fig:fit} shows our analytic fit to the optical and mid-infrared lightcurves. 

We fit the existing pre-outburst and outburst optical photometry 
with a sigmoid function, assuming the peak brightness occurs 
at the last measured epoch, which is defined as $t=0$.
The sigmoid is described by $L/(1 + e^{(-k\times (t-t_o)/days)})$  
and our best-fit parameter values are: 
$t_o=-522.9\pm 0.9$ (meaning that the outburst starts 2$\times t_o$=1046 days before its peak), 
$k=-0.00728\pm 0.00043$ (where the $1/k$ value represents a timescale 
for the rise, in this case a 137.4 day e-folding), 
and $L=3.541\pm 0.002$ (representating the full amplitude of the rise).
We note that it is not yet clear whether \gbpi\ has reached its peak brightness.
If it continues to rise, all of the numbers quoted above would increase.

Applying the same methodology to the infrared data, we fit the first and second
plateaus in the {\it NEOWISE} lightcurve separately. 
We find that the first rise has a 71 day e-folding time, 
taking 2$\times t_o$= 593 days to rise 2.2 mag,
while the second rise has a 149 day e-folding time, 
taking 2$\times t_o$= 998 days to rise another 1.1 mag.
The second rise in the {\it NEOWISE} data appears to correpond to
the brightening in the {\it Gaia} data.
Again, it is not clear whether the second infrared rise has reached its peak,
so the numbers above could be lower limits.

The amplitude of the photometric brightening corresponds to
a luminosity increase by a factor of 25, from the estimated 0.3 $L_\odot$
for the progenitor T Tauri star, to 7.5 $L_\odot$ in outburst.

Figure~\ref{fig:color} illustrates the optical and mid-infrared color changes
that have been observed.
Although there is little optical color data available for \gbpi\
in the outburst phase (Table~\ref{tab:optphot}), compared to pre-outburst colors
from IPHaS and PanSTARRS, the source is clearly bluer. 
After accounting for AB to Vega magnitude transformations, the blue-ing 
is about 1.2 mag in r-i color and 0.8 mag in g-r color. 
There should be a full g-r color time series available from ZTF when
the first data release occurs.
In addition, a future data release from Gaia should contain $BP$ and $RP$
measurements at the same epochs as the $G$ measurements.

In the infrared, the {\it Spitzer} measurements show 
no infrared color change between the 2004 and 2014 data.
However, the first {\it NEOWISE} measurement seems to indicate
a redder [3.4]-[4.6] color (a.k.a. W1-W2) compared to 
the historical {\it Spitzer} [3.6]-[4.5] color (a.k.a. I1-I2), by $\sim$0.45 mag. 
Color terms between the {\it Spitzer} and {\it WISE} filter systems can not
explain the amplitude of the apparent color change; green points 
in Figure~\ref{fig:color} show the {\it Spitzer} photometry corrected to 
the {\it NEOWISE} system using an empirically derived 
relation\footnote{$(W1-W2) = 1.62\times(I1-I2) - 0.04$ mag, with rms=0.24 mag}.
The {\it NEOWISE} outburst photometry is initally redder,
but as the optical and second-stage-infrared burst proceeds, 
the mid-infrared color is observed to become bluer, by $\sim$0.4 mag.  
The fitted color-magnitude slope is $W1 = 2.07\times(W1-W2)+10.70$ mag
with rms=0.13 mag.  This slope is inconsistent with the clearing of
extinction.

The evidence for significant and substantial blue-ing during the outburst, 
in both optical and infrared colors,
points to a dramatic heating event in \gbpi.

\subsection{SPECTROSCOPIC FOLLOW-UP DATA AND ANALYSIS}

\begin{figure}
\includegraphics[width=0.5\textwidth]{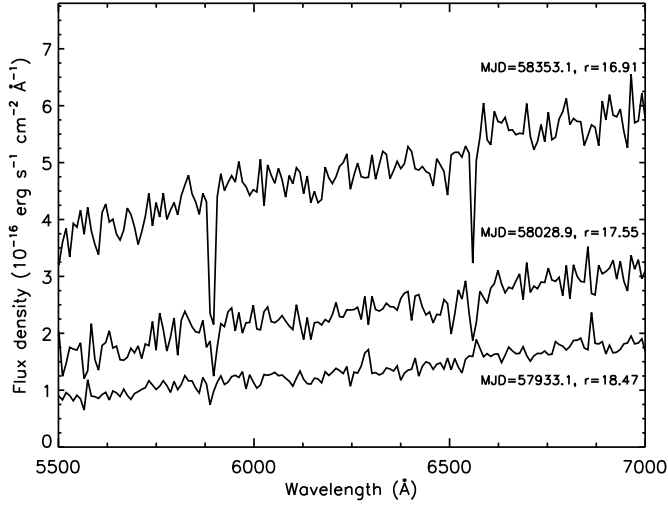}
\caption{
Portion of the Liverpool Telescope spectra of \gbpi\ at R=350.
From bottom to top, representing three different nights during 
the lightcurve rise, NaD and $H\alpha$ absorption appears and strengthens.
}
\label{fig:optspec_lt}
\end{figure}

During the lightcurve rise, the Liverpool Telescope was used to obtain 
optical spectra using the low-resolution spectrograph SPRAT (R=350). 
The observations were carried out during the same nights 
as the photometric monitoring reported above.
The data are displayed in Figure~\ref{fig:optspec_lt}.

The first low resolution spectrum, obtained when the source was $\sim$2 mag brighter than quiescence, 
shows little in terms of emission or absorption. But as the source brightens
to more than $\sim$3 mag above quiescence, distinct absorption features in
NaD and $H\alpha$ appear, and strengthen over time.

\begin{figure}
\includegraphics[width=0.7\textwidth,angle=-90]{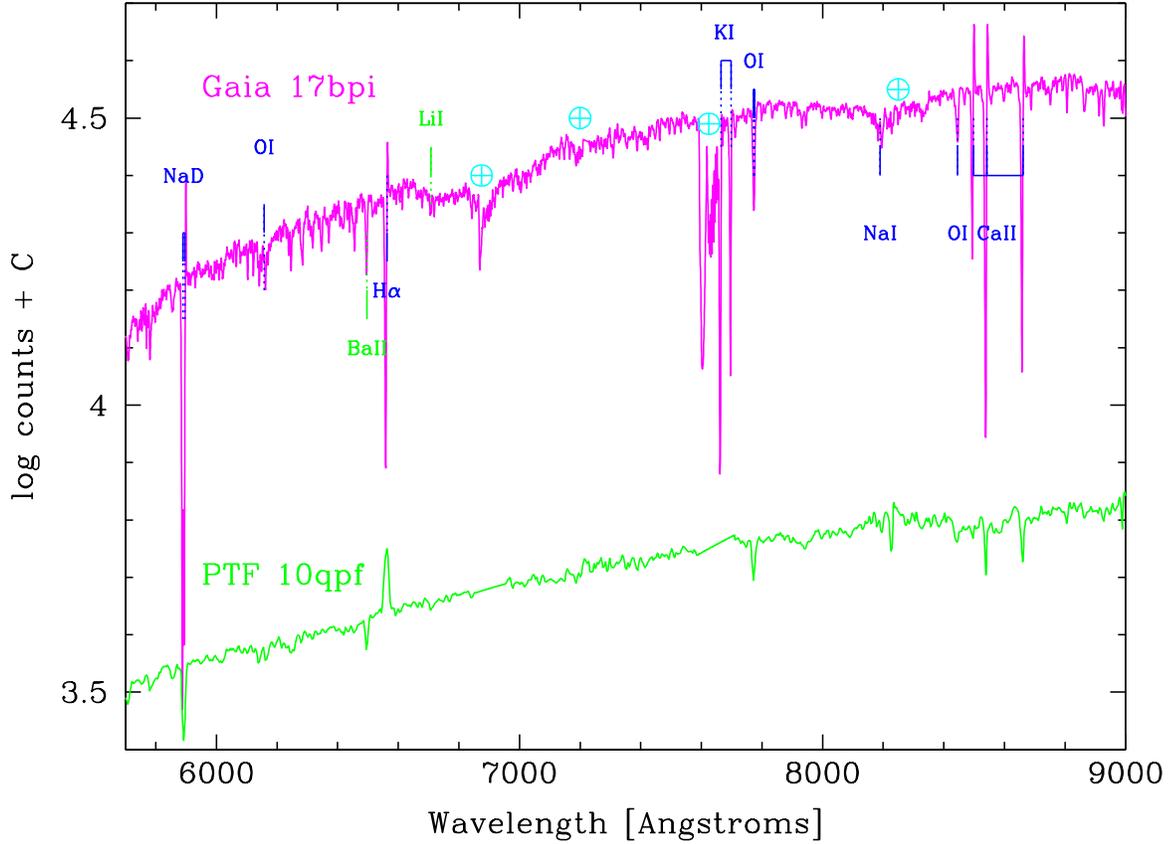}
\caption{
Portion of the Keck/DEIMOS spectrum of \gbpi\ at R=2200.
Prominent spectral lines are labelled, with the $\Earth$ symbol indicating
regions of significant telluric contamination.  The spectrum is consistent
with that of a GK-type photosphere, and comparable to the spectrum
of PTF 10qpf = V2493 Cyg (formerly known as LkH$\alpha$188/G4 = HBC 722),
which was observed at a similar time in its outburst, 
though at lower resolution and with the telluric-contaminated regions 
interpolated over.
}
\label{fig:optspec_keck}
\end{figure}

\begin{figure}
\includegraphics[width=0.50\textwidth,trim={0.75cm 0 0.75cm 0},clip]{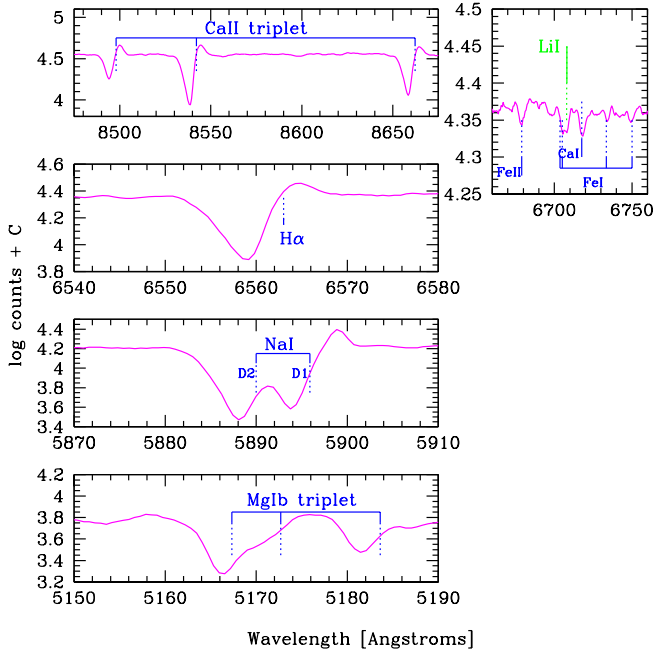}
\vskip-0.5truein
\caption{
Zoom-in on portions of Figure~\ref{fig:optspec_keck} illustrating
the P-Cygni nature of the NaD, H$\alpha$, and \ion{Ca}{2} 
line profiles, as well as the blueshifted absorption in 
the Mgb triplet lines - all signatures of a strong wind.  
\gbpi\ clearly exhibits \ion{Li}{1} absorption, 
which appears to also have a blueshifted component 
that exceeds possible contributions from weak \ion{Fe}{1} lines.
}
\label{fig:profiles}
\end{figure}

The Keck/DEIMOS \citep{faber2003} spectrograph was then used on 2018, September 10 UT 
by E.N. Kirby to observe \gbpi\ in a 900 sec integration.  The 600ZD grating provided
dispersion of 0.65 \AA/pix, and led to a resolution of 3.3 \AA\ per 0.7" slit, or
$R\approx 2200$ over 4630 to 9865 \AA, with a small gap from 7211-7229 \AA. 

The DEIMOS spectrum (Figure~\ref{fig:optspec_keck}) exhibits numerous strong 
P-Cygni-type blueshifted absorption lines, with accompanying weak redshifted emission 
components.
These profiles are seen in NaD, H$\alpha$, and the \ion{Ca}{2} triplet as
highlighted in Figure~\ref{fig:profiles}.  Blueshifted absorption,
but non-P-Cygni profiles are presented by the Mgb triplet, \ion{K}{1}, and
\ion{O}{1}.  All of the above lines indicate a strong wind 
with velocity several hundred km/s.  The terminal velocity of the
H$\alpha$ profile, for example, is approximately $-500$ km/s.
\ion{Li}{1} absorption is also present 
(Figure~\ref{fig:profiles}) and has some evidence for a wind component as well. 
The \ion{Li}{1} equivalent width $W_\lambda$ is 0.47 \AA, 
and is notable relative to the nearby \ion{Ca}{1} 6717 \AA\
line with $W_\lambda$=0.39 \AA, though there could be minor contamination
from weak \ion{Fe}{1} lines at 6703.6 and 6705.1 \AA.  

The H$\alpha$ profile has $W_\lambda$=9.2 \AA\ in blueshifted absorption 
with a double-trough, and then a redshifted emission component with 
$W_\lambda$=-1.1 \AA.
In H$\beta$ there is $W_\lambda$=5.1 \AA\ strength in blueshifted absorption 
and no emission component.
The \ion{Ca}{2} 8542 \AA\ lines has $W_\lambda$=5.2 \AA\ in its blueshifted 
absorption and $W_\lambda$=-1.3 \AA\ in redshifted emission.

In addition to the lines indicating activity and youth, the optical spectrum
of \gbpi\ also has numerous neutral species absorption lines that are
typical of GK type spectra, e.g. \ion{Fe}{1}, \ion{Mg}{1}, \ion{Ca}{1}.
Also like many FU Ori stars, \gbpi\ has a strong feature at 6497 \AA\ 
that is associated with a \ion{Ba}{2}/\ion{Ca}{1}/\ion{Fe}{1} blend.

\begin{figure}
\includegraphics[width=0.9\textwidth,trim={0.75cm 0 0.75cm 0},clip]{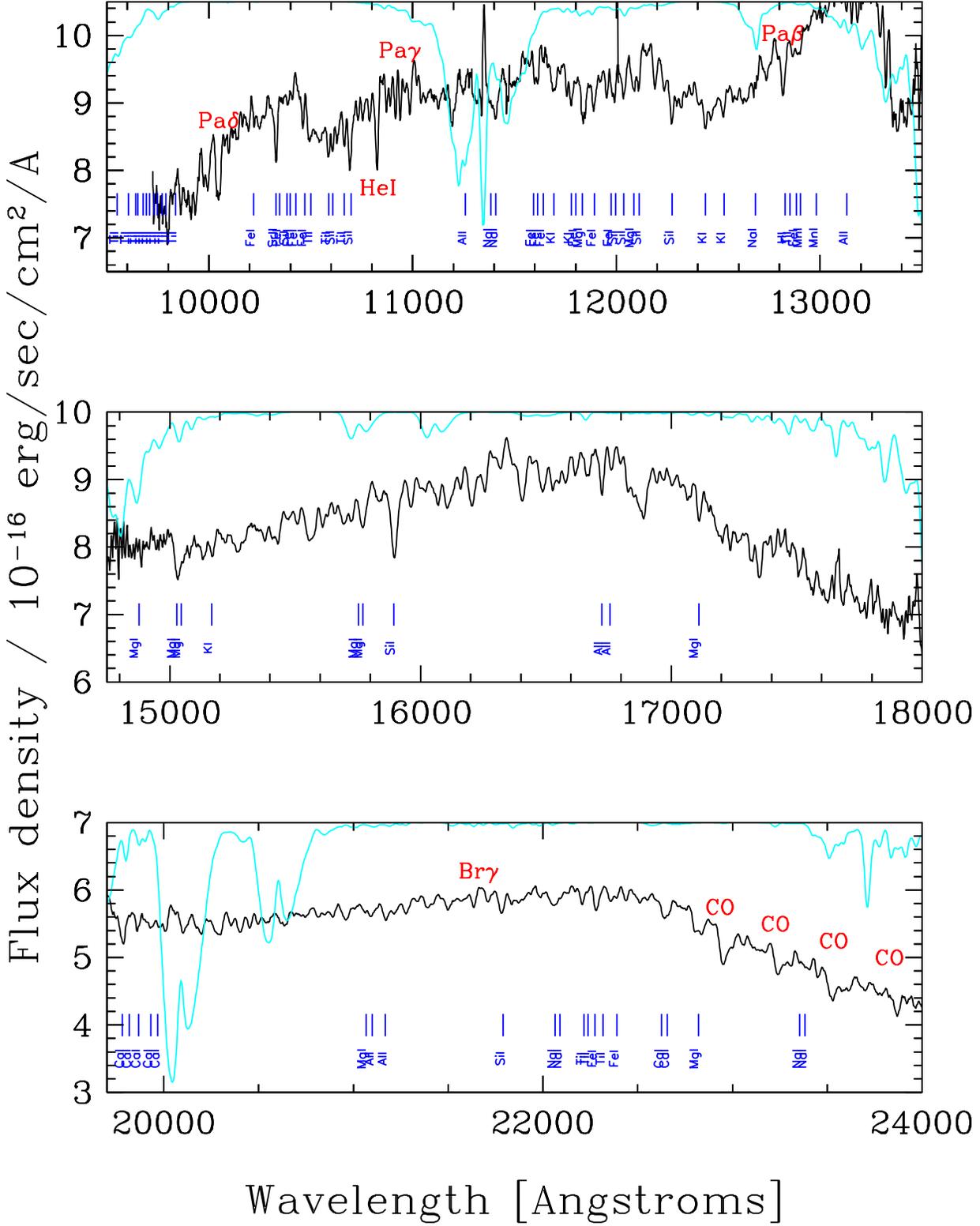}
\caption{
Portion of the Palomar/TripleSpec spectrum of \gbpi\ at R=2700.
Prominent spectral lines are labelled.  
The cyan line is a model atmospheric transmission spectrum 
plotted on a vertical scale from 0-100\%
and indicates regions where the telluric correction applied to the data 
is large and possibly uncertain.
The \gbpi\ spectrum is similar to that of an M-type photosphere, mainly due to the
prominent $H_2O$ broad depressions in the H-band and K-band,
and the $^{12}$CO (2-0) bandhead absorption in the K-band. 
The spectrum agrees well with the set of FU Ori objects displayed
in Figure 3 of \cite{connelley2018}.  
}
\label{fig:irspec}
\end{figure}

The Palomar 200" telescope and TripleSpec \citep{herter2008} instrument were used
on 2018, September 25 UT to observe \gbpi\ in the YJHK bands.  A total of
1 hour of integration was accumulated from 3 sets of 4 (A-B-B-A) nods. 
Data were
reduced using a version of the XSpexTool developed by M. Cushing and W. Vacca
adapted for TripleSpec at Palomar.
The spectrum
has resolution $R=2700$ and is illustrated in Figure~\ref{fig:irspec}.

The classic FU Ori signatures of $^{12}CO$ absorption in the 2.3 micron region 
and prominent $H_2O$ absorption in the K-band and H-band
are exhibited, as are absorption features due to various atomic lines. 
Notably \ion{He}{1} 10,830 \AA\ and several \ion{H}{1} Pa lines are apparent. 
Although not apparent at our resolution, these lines likely have
P Cyg type wind profiles similar to those exhibited by the optical region
and \ion{H}{1} Balmer and \ion{Ca}{2} lines.
We refer to the FU Ori spectral atlas of \cite{connelley2018} 
and specifically to their Figure 3 for a set of comparison objects. 

\section{DISCUSSION AND CONCLUSIONS}

There is wide diversity among members of the FU Ori class regarding outburst rise times and amplitudes.
\gbpi\ has exhibited a 3.5 mag rise in the optical over 1046 days, with a 137.4 day e-folding time,
and an accompanying $>$3 mag brightness increase in the mid-infrared. 
The infrared brightening took place in two stages, the first beginning
approximately 1.5 years before the optical brightening 
(late in 2014 and into 2015, compared to in 2017), and the second seemingly coincident with
the optical brightening.  
Source colors show both mid-infrared and optical blue-ing during the lightcurve rise.  

The unifying elements of the FU Ori class are the distinctive spectroscopic features, notably 
a spectral change from hotter to cooler with increasing wavelength through the optical
and near-infrared, as well as the appearance of strong wind signatures that are a consequence of
the onset of rapid accretion.  \gbpi\ exhibits a GK-type absorption spectrum in the optical, 
and an M-type absorption spectrum in the infrared, as is typical of the FU Ori class.
While the spectra are similar in the different wavelength ranges to these 
single-temperature spectral classes, they are not exact matches 
given the complexity of the ``photospheres" of members of the FU Ori class.
\gbpi\ also shows spectroscopic signatures of strong wind/outflow, and it displays 
the \ion{Li}{1} 6707 \AA\ signature of youth.
These characteristics are also consistent with an FU Ori classification.
 
With the data in hand to date, \gbpi\ appears to meet the photometric and spectroscopic critera 
of a bona fide FU Ori source.  The source thus joins 
V900 Mon \citep{thommes2011,reipurth2012}, 
V960 Mon \citep{maehara2014,hillenbrand2014}, 
[CTF93]216-2 = V2775 Ori \citep{caratti2011}, and 
HBC 722 = V2493 Cyg \citep{semkov2010,miller2011} 
as the latest entries over the past ten years in this still-rare category.
Fewer than 13 of the known $\sim$25 FU Ori objects have been observed to undergo their 
dramatic brightness increases, with the rest classified as such only after the fact, 
e.g.  V582 Aur \citep{samus2009,munari2009} was relatively recently recognized as a 
member of the class, in 2009, but likely outburst in the mid-1980's \citep{semkov2013}.

\gbpi\ is unique among FU Ori outbursts in having its photometric brightening
detected at both optical and mid-infrared wavelengths.  The burst appears to have
started in the infrared, consistent with disk models 
that predict instabilities in the inner 0.5-1 AU of 
protostellar and T Tauri accretion disks as the origin of FU Ori events. 

\section{ACKNOWLEDGMENTS}
We acknowledge ESA Gaia, DPAC and the Photometric Science Alerts Team, 
as well as NASA's NEOWISE team. 
We are extremely grateful to Evan Kirby for obtaining and reducing to 1D format
the Keck/DEIMOS spectrum presented here. 
We are also grateful to the Palomar Observatory staff for their real-time
assistance in obtaining the TripleSpec spectrum presented here. 
Conversations with Kevin Burdge and Sean Carey 
about various photometric survey data sets were beneficial in our work.
The contributions of CCP and TN were funded by a Leverhulme Trust Research Project Grant, and of SM through a Science and Technology Facilities Council studentship.

\facility{Gaia, LT:IO:O, LT:IO:I, BO, Hale:DBSP, Hale:TSPEC, Keck:I:DEIMOS, IPHaS, PanSTARRS, 2MASS, Spitzer, Herschel, WISE, NEOWISE, IRSA}

\end{document}